\documentclass[twocolumn]{article}

\usepackage{graphics}

\usepackage{amsmath,amssymb}
\usepackage{graphicx}
\usepackage{authblk}

\title{Frequency doubled telecom fiber laser for a cold atom interferometer using optical lattices}

\author[1]{Fabien Theron}
\author[1]{Yannick Bidel}
\author[1]{Emily Dieu}
\author[1]{Nassim Zahzam}
\author[1,2]{Malo Cadoret}
\author[1]{Alexandre Bresson}
\affil[1]{ONERA - The French Aerospace Lab, BP 80100, 91123 Palaiseau Cedex, France}
\affil[2] {Laboratoire Commun de M\'etrologie, CNAM, 61 rue du Landy, 93210 La Plaine Saint-Denis, France}

\begin{document}
\maketitle

\begin{abstract}
A compact and robust laser system, based on a frequency-doubled telecom laser, providing all the lasers needed for a rubidium cold atom interferometer using optical lattices is presented. Thanks to an optical switch at 1.5 $\mu$m and a dual-wavelength second harmonic generation system, only one laser amplifier is needed for all the laser system. Our system delivers at 780 nm a power of 900 mW with a detuning of 110 GHz for the optical lattice and a power of 650 mW with an adjustable detuning between 0 and -1 GHz for the laser cooling, the detection and the Raman transitions.\\
\end{abstract}

\section{Introduction}

Cold atom interferometers \cite{AI school} have demonstrated to be very accurate and sensitive sensors to measure gravity acceleration \cite{gravi Peters,gravi Syrte,gravi Chinois}, gravity gradient \cite{gradio Kasevich,gradio Tino} and rotation \cite{gyro Kasevich,gyro Syrte}. Now, more and more experiments are using optical lattices \cite{optical lattice} to improve the performance of these sensors. Indeed, optical lattices allow to control in an extremely precise way the momentum of the atoms and thus its movements. It is therefore possible to levitate atoms with a perfectly controlled force and thus to build compact gravimeter with levitating atoms \cite{renee,clade}. It is also possible to increase the momentum transfer in an atomic splitter and thus to increase the sensitivity of the sensor \cite{LMT Kasevich,LMT Muller,LMT Robins}. Optical lattices can also be used to impart precisely a well controlled velocity to the atoms which is very important for gyroscope or large atomic fountain experiment \cite{Large fountain}. Finally optical lattices can be used to separate and transport a cold atom cloud in order to build a sensor with spatially separated clouds of cold atoms and thus to measure gravity gradient \cite{gradio spatial}. The disadvantage of using optical lattices in atom interferometry sensors is that one needs another laser generally detuned around 100 GHz from the atomic transition with a power of at least few hundred mW. Until now, the optical lattices were implemented with an independent laser different from the one used for the atom interferometer experiment. The whole laser system for the sensor is thus more complicated, larger and less robust. This is a real inconvenient if one builds a compact and robust sensor \cite{girafe} for geophysics, navigation or space application. Many developments \cite{carraz,schmidt,cheinet} have been done over the last few years to achieve compact and robust laser systems for atom interferometers but none of them were addressing optical lattices.

In this article, we present a laser system which generates the lasers for a cold atom interferometer and optical lattices with only one laser amplifier which is usually the less reliable part. Our laser system is based on a frequency doubled telecom fiber bench which addresses rubidium atoms ($\lambda=780\,$nm) and is derived from the laser system described in the reference \cite{renee} in which two independent laser systems composed of a laser source at 1.5 $\mu$m, an EDFA (Erbium Dopped Fiber Amplifier) and a second harmonic generation bench were implemented: one for the cold atom interferometer and one for the optical lattices. The laser system presented in this article combines the two laser sources at 1.5 $\mu$m thanks to an optical switch and thus allows to use only one EDFA. This trick is possible because the optical lattice laser and the cold atom interferometer laser are not needed at the same time. In the first part of this article, we will present the general architecture of our laser system then we will describe the dual wavelength second harmonic generation and finally we will present the lock of the optical switch.

\begin{figure*}[ht]
	\begin{center}
	\centerline{\scalebox{0.6}{\includegraphics{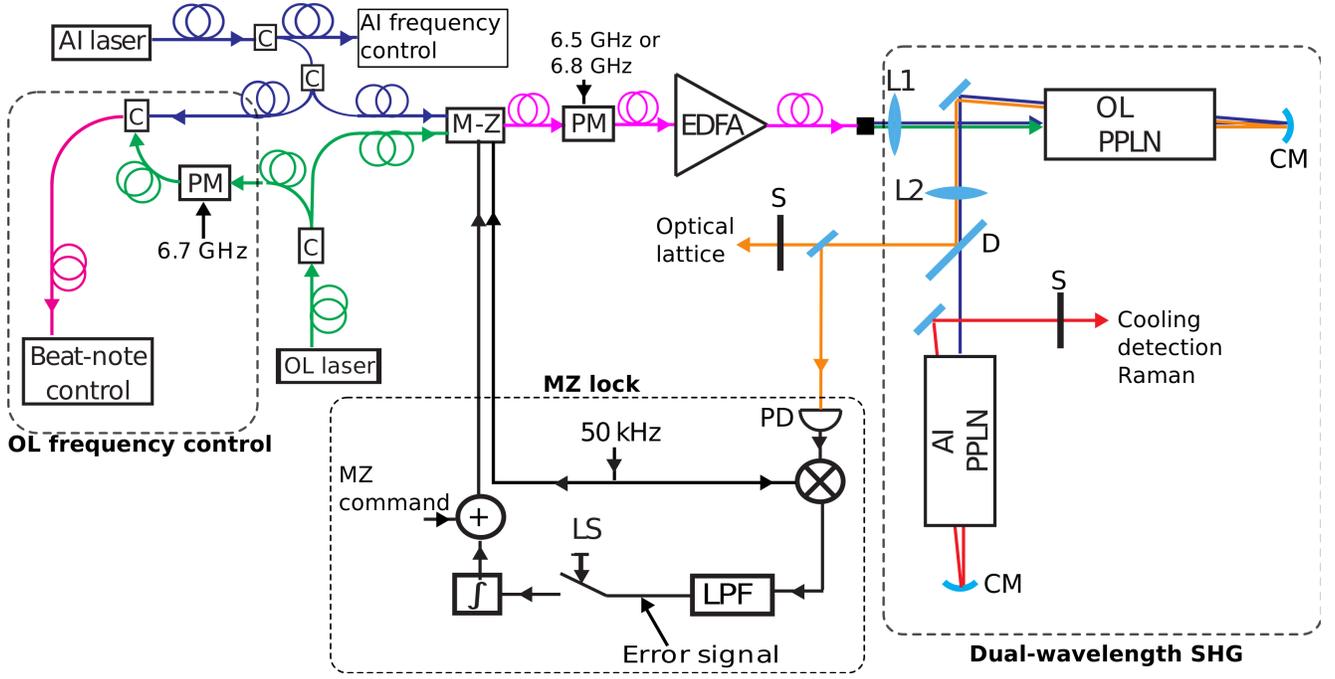}}}
	\end{center}
\caption{Diagram of the laser system and the Mach-Zehnder lock. AI: Atom Interferometry, OL: Optical Lattice, C: coupler, PM: Phase Modulator, M-Z: Mach-Zehnder, EDFA: Erbium Doped Fiber Amplifier, L: Lens, PPLN: Periodically-Poled Lithium Niobate, D: Dichroic mirror, CM: gold Concave Mirror, S : Shutter, PD: PhotoDiode, LPF: Low Pass Filter, LS: Lock Switch.}
\label{laser}
\end{figure*}

\section{General description of the laser system}

Our system generating the lasers for the cold atom interferometer and the optical lattices is described on Fig. \ref{laser}. The atom interferometer laser source (AI laser) which provides the frequencies for laser cooling and trapping, detection and Raman transition is an erbium fiber DFB laser at 1.5 $\mu$m (output power: 20 mW, linewidth 2 kHz). This laser source is locked relative to the rubidium transitions with a detuning which can be adjusted from 0 to -1 GHz. The repumper and the second Raman frequency are generated thanks to a phase modulator fed by a radio frequency at 6.5 GHz or 6.8 GHz. This part of the laser is described in details in reference \cite{AI}. The optical lattice laser source (OL laser) is a DFB laser diode at 1.5 $\mu$m (Avanex DFB, output power: 10 mW, linewidth 1 MHz). The OL laser is detuned by 55 GHz at 1.5 $\mu$m relative to the AI laser which gives a detuning of 110 GHz at 780 nm. The detuning of the OL laser is controlled thanks to a beat-note at 1.5 $\mu$m between the two lasers. In order to have a beat-note at lower frequency easily measurable, a phase modulator fed by a radio frequency at 6.7 GHz is put on the optical lattice path and generates sidebands separated by 6.7 GHz. Then, a beat note at 3.1 GHz is measured between the AI laser and the 7$^{th}$ sideband of the modulated OL laser which allows to control the frequency of the OL laser.

The AI laser and the OL laser are combined at 1.5 $\mu$m thanks to an electro-optical modulator which acts like a continuous optical switch between each laser. This component is a wave-guided Mach-Zehnder interferometer (MZ) from EO Space with an extinction of 30 dB and a period of 5 V. Because the path difference in the MZ is close to zero, interference signals of OL laser and AI laser are in phase opposition and the output of the MZ can be described by Eq. \ref{MZ}. Thus changing the MZ voltage command V allows to select the AI laser or the OL laser at the output.

\begin{equation}
	I_{out} = I_{AI} \times \cos^{2}(\alpha V) + I_{OL} \times \sin^{2}(\alpha V),~ \alpha \in \Re
	\label{MZ}
\end{equation}

Then, the laser goes through a 5 W Erbium Doped Fiber Amplifier (EDFA). In order to avoid damaging the EDFA, one must make sure to have a constant power at its input. The power of both laser sources have therefore been adjusted to have the same power at the output of the MZ regardless of the MZ command. Finally, the laser is sent to the dual-wavelength second harmonic generation bench.

\section{Dual-wavelength frequency second harmonic generation}

\begin{figure}[!ht]
\centerline{\scalebox{0.7}{\includegraphics{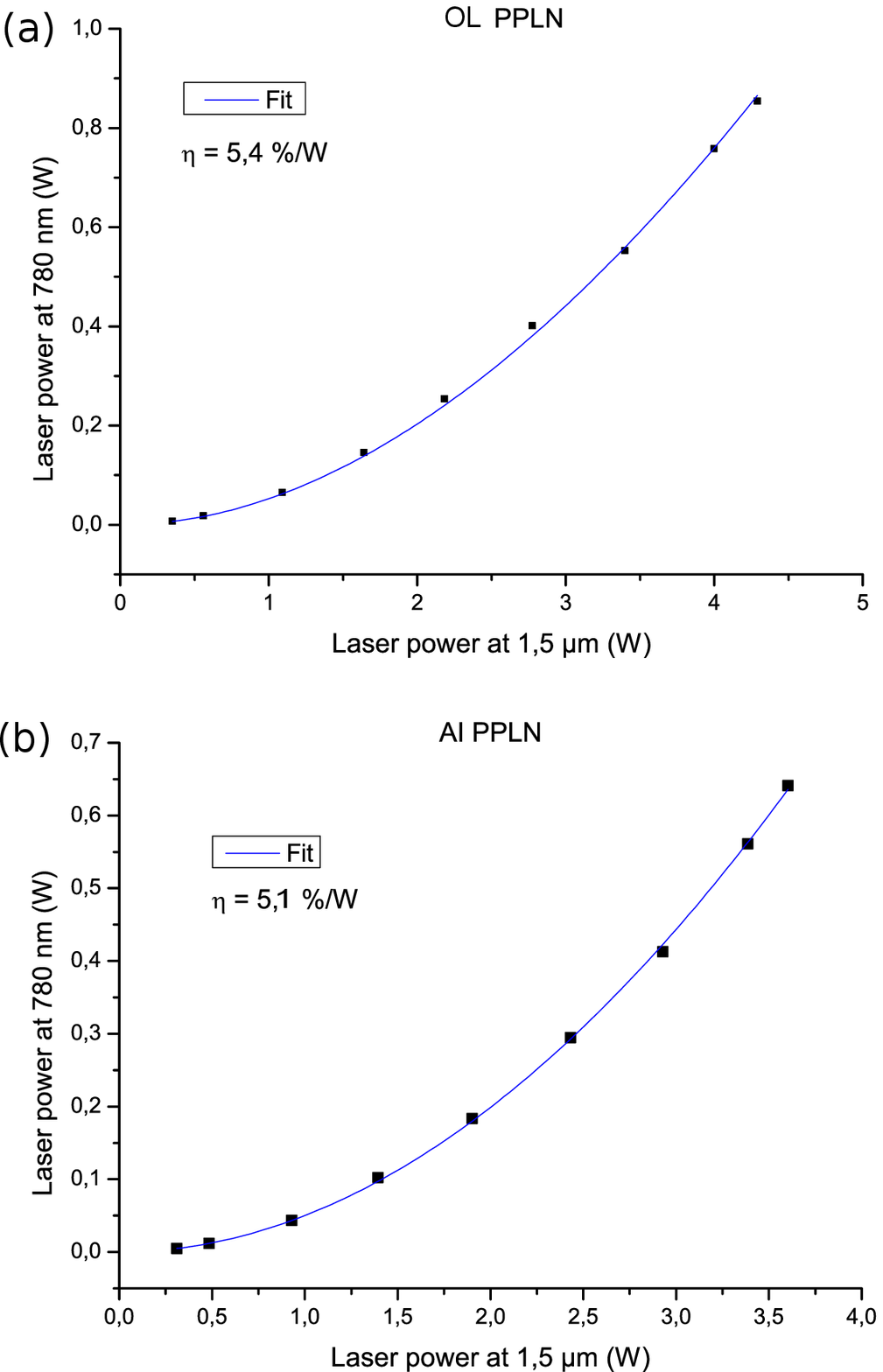}}}
\caption{Second harmonic generation at 780 nm of the laser. On top (a), doubling efficiency of the OL laser, with the OL PPLN, depending pump power. On bottom (b), doubling efficiency of the AI laser, with the AI PPLN, depending pump power.}
\label{doubling}
\end{figure}

The dual-wavelength second harmonic generation configuration that we have implemented is similar to the one described in reference \cite{dual}. It consists of two bulk PPLN crystals, in double pass configuration (Fig. \ref{laser}).

At the output of the EDFA fiber, the laser is focused in a first PPLN crystal thanks to a 75 mm focal length lens (L1), and the double pass configuration is realized thanks to a 10 cm curvature radius gold mirror (CM). At the center of the crystal, the waist of the laser is 23 $\mu$m. This first PPLN is phase matched for OL frequencies (T$_{OL}$ = 133.45 \textordmasculine C) and allows to get an OL laser power of 900 mW at 780 nm with an input power at 1.5 $\mu$m of 4.3 W. The measurement of the power at 780 nm versus the power at 1.5 $\mu$m is given on Fig. \ref{doubling}. The experimental data are fitted by Eq. \ref{conversion} which takes into account the pump depletion \cite{SHG}. 
\begin{equation}
P_{780}\, = P_{1.5}\,\tanh^{2}(\sqrt{\eta P_{1.5}})
\label{conversion}
\end{equation}
The fit gives a conversion efficiency of $\eta$ = 5.4 $\%/W$.

The dichroic mirror (D) transmits the AI laser at 1.5 $\mu$m to the second PPLN, and reflects the OL laser at 780 nm. The AI laser at 1.5 $\mu$m is focused in the second PPLN crystal thanks to a 60 mm focal length lens (L2) in 2f-2f design, in order to get the same waist for both crystals. The double pass configuration in the crystal is realized thanks to a 10 cm curvature radius gold mirror. The second PPLN is phase matched for AI frequencies (T$_{AI}$ = 134 \textordmasculine C) and allows to get an AI laser power of 650 mW at 780 nm with an input power at 1.5 $\mu$m of 3.5 W. For this crystal, the fit of the power at 780 nm versus the power at 1.5 $\mu m$ gives a conversion efficiency of $\eta$ = 5.1 $\%/W$ (see Fig. \ref{doubling}).

\begin{figure}[!ht]
\centerline{\scalebox{0.45}{\includegraphics{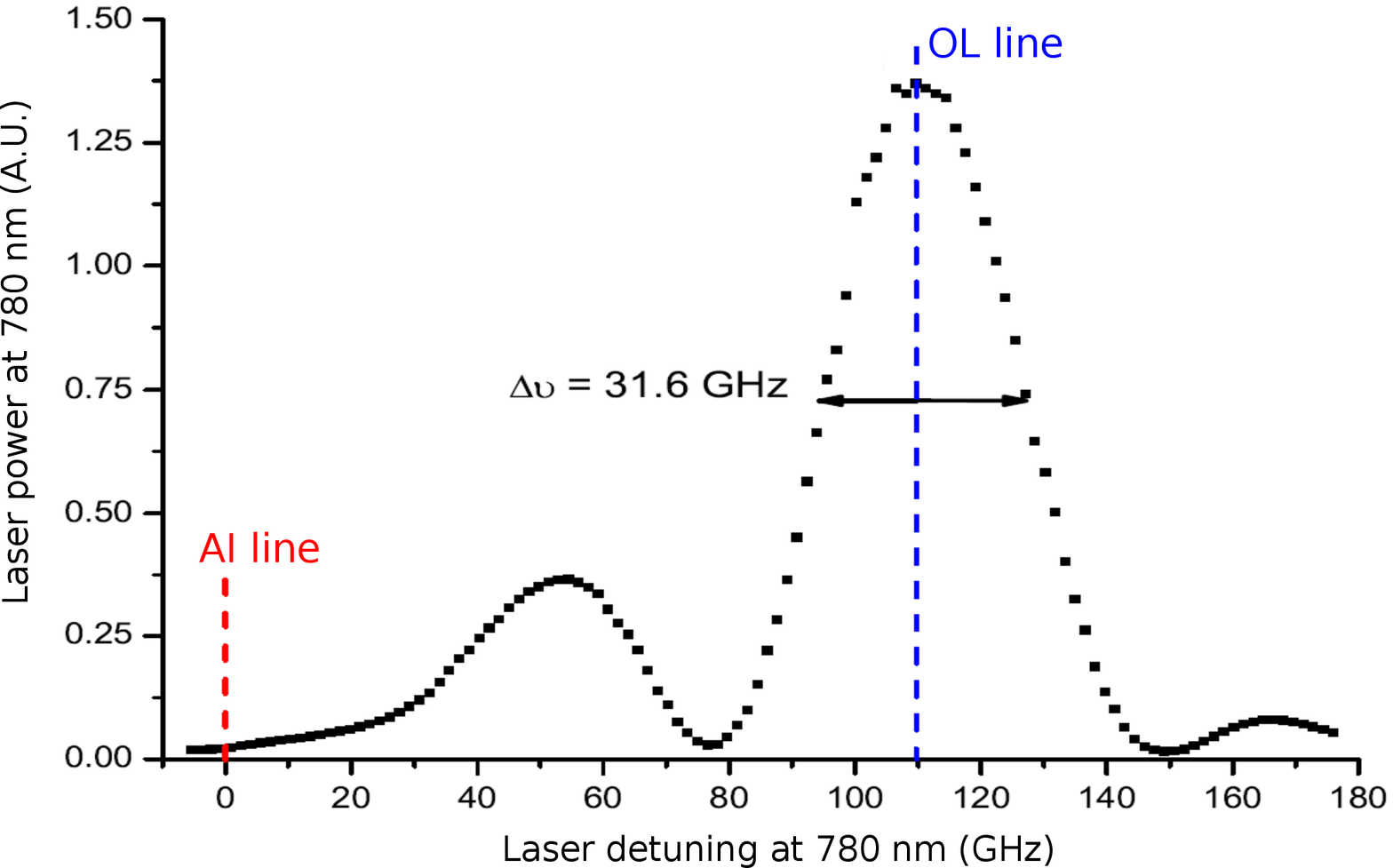}}}
\caption{Second harmonic generation at 780 nm in the PPLN phase matched for OL conversion, with 1.5 $\mu$m laser at 4.5 W. OL line in blue and AI line in red.}
\label{conv}
\end{figure}

\section {Effect of the finite extinction of the optical switch}

Due to the finite extinction of the optical switch, the OL and the AI laser beams will be contaminated by a small fraction of light at the frequency of the wrong laser and at the sum frequency of the 1.5 $\mu$m AI laser and the 1.5 $\mu$m OL laser. The parasite laser frequency which has the biggest impact on the experiment is the AI frequency in the OL laser beam because it is close to resonance and could induce extra spontaneous emission in the optical lattice. It is thus important to evaluate the relative power of this parasite laser frequency and to estimate its effects on the spontaneous emission rate. We measure an extinction of the optical switch equal to $\eta_{MZ}=1\times\,10^{-3}$. By measuring the power at 780 nm versus the frequency of the laser for the OL crystal (see Fig. \ref{conv}), one finds that the ratio between the conversion efficiency at the AI frequency  and the conversion efficiency at OL frequency is equal to $\eta_{conv} = 1.5 \times$10$^{-2}$. The relative power at AI frequency in the OL laser beams is equal to $\epsilon_{AI}=\eta_{MZ}^2\times\eta_{conv}=1.5\times 10^{-8}$. During the optical lattice stage, the AI can be detuned by $\Delta_{AI}=1$ GHz compared to the atomic resonance. Thus, the AI parasite laser line causes a spontaneous emission rate $\epsilon_{AI}\times\frac{\Delta_{OL}^2}{\Delta_{AI}^2}=1.8\times 10^{-4}$ less than the spontaneous emission rate caused at the OL laser line. We can conclude that the effect of AI parasite frequency in the OL beam will be completely negligible.

\section{Lock of the Mach-Zehnder optical switch}

The MZ switch presents a long term drift on hours time scales which results in a drift of the command voltages needed to select AI and OL frequencies. A continuous correction of the MZ voltage command has thus been implemented by modulating the voltage of the MZ at 50 kHz and by detecting this modulation on the power at 780 nm of the OL beam (Fig. \ref{laser}). For that purpose, a part of the OL beam is sent to a photodiode. Then the signal is amplified, demodulated at 50 kHz and low pass filtered. The error signal obtained at this point is integrated, summed to the electric command which switch from AI to OL and sent to the MZ input. Because the error signal is only available when the OL frequency is selected, an electronic switch (LS) has been inserted before the integrator in order to activate the lock only during the optical lattice stage. During the AI stage, the output of the integrator will thus stay at a constant value. This non correction of the drift of the MZ during the AI stage is totally acceptable because the duration of this step is very short ($\sim 1$ s) compared to the time scale of the drift of the MZ (few hours).

\section{Conclusion}

We have developed a compact and robust laser system providing all the lasers needed for a cold atom interferometry sensor using optical lattices. The association of a fibered Mach-Zehnder electro-optical switch and a dual-wavelength second harmonic generation unit has allowed us to use only one laser amplifier and thus lead to a laser system which is more reliable, more compact and with a lower electrical consumption.  We have checked that the finite extinction of the switch does not cause additional spontaneous emission in the optical lattice. A power of 650 mW for the cold atom interferometer and 900 mW for the optical lattice have been obtained. These laser powers can be increased by using a 10 W laser amplifier which is a component commercially available. By taking parameters of the dual-wavelength architecture and using Eq. \ref{conversion}, one obtains an OL laser power of 3 W, and an AI laser power of 2 W. The laser system presented in this article is addressing transportable and onboard atomic sensor with enhanced performance thanks to optical lattice technology and can have applications in space technology, geophysics or navigation. \\

\textbf{\scriptsize{ACKNOWLEDGEMENTS}} \, We thank F. Nez, from the Laboratoire Kastler
Brossel (LKB), for his help on the project. We acknowledge funding support from the Direction Scientifique G\'en\'erale of ONERA and the Direction G\'en\'erale de l'Armement (DGA).

\end{document}